# Radio Interface for High Data Rate Wireless Sensor Networks

J. Hénaut, A. Lecointre, D. Dragomirescu, R. Plana
LAAS-CNRS, University of Toulouse, 8 Avenue de Colonel Roche, 31077 Toulouse, France
Email: {jhenaut, alecoint, daniela, plana}@laas.fr

**ABSTRACT**

**This paper gives an overview of radio interfaces devoted for high data rate Wireless Sensor Networks. Four aerospace applications of WSN are presented to underline the importance of achieving high data rate. Then, two modulation schemes by which High Data Rate can be achieved are compared : Multi carrier approaches, represented by the popular Orthogonal Frequency Division Multiplexing (OFDM) and Single carrier methods, represented by Single Carrier Frequency division Equalization and its application for multiple access Single Carrier Frequency division multiple Access (SC-FDMA). SC-FDMA, with a very low Peak Average Power Ratio (PAPR), is as strong alternative to the OFDM scheme for highly power constraint application. The Chosen radio interface will be, finally, tested by a model based design approach based on Simulink and FPGA realization.**

## 1. INTRODUCTION

MIT's Technology Review Magazine called Wireless Sensor Networks (WSN) « one of the ten technologies that will change the world ». No wonder that many papers and studies are done on the subject. Sensor nodes perform several functions such as: sensing physical parameters of the environment, locally processing raw data to extract characteristic features of interest, momentarily storing this information, and transmitting it to its neighbors through a wireless link. To perform all these functions, sensor nodes include many sub-systems which can be integrated into a single system-on-a-chip (SoC) in order to minimize power consumption and reduce the cost. Since sensor networks are self powered, energy limitation introduces severe constraints for the design of wireless sensor nodes. Today, at the physical layer, radio interface remains one of the bottlenecks to implement truly low-power WSN. One of the most popular solutions is to develop WSN architectures featuring low data rates. As the node only needs to communicate from time to time, one can introduce a sleeping mode to minimize the power consumption of the radio interface and therefore, that of the entire node. Unfortunately, for high data rate applications, designers have to rethink the entire system architecture.

The aim of this paper is to define a radio interface such as to enable high data rate wireless Sensor Networks. It will be shown that dramatic advances are possible if more attention is paid to the Radio Interface (RI).

It has to be outlined that considering a radio interface suitable for all kinds of applications is a rather complex problem as it will involve many issues competing between each other. For example, modulation choice is always a tradeoff between power consumption, Bit Error Rate, and data rate.

For several reasons, we chose, to investigate WSN in the 5 GHz range. Other frequency range like 60 GHz has not been retained at this stage for technology maturity and cost. Frequencies lower than 5 GHz is not studied, for size issues.

This paper is organized into seven sections. Section 2 introduces three kinds of WSN applications for aerospace. Once the needs are clearly indentified, three types of radio interface are investigated: Narrowband modulation (Section 3), OFDM based transmission (Section 4) and SC-FDE/SC-FDMA (Section 5). Section 6, proposes design flow for WSN applications. Finally conclusions are outlined in the last section.

## 2. WSN AEROSPACE APPLICATIONS

Wired links tend to be challenging for aircraft industry due to excessive weight and complexity. The same problems can concern all industry branches, i.e. car manufacturing. As a matter of fact, a wide number of electronic systems need to communicate inside a car. The bus load can be significantly decreased if the sensors submit the respective signals wirelessly. Four specific applications of wireless sensor networks could prove to be useful to reduce both weight and complexity.

The first application that could benefit from wireless connection is referred to as "In Flight Entertainment" (IFE). Airlines companies wish to offer wider distraction opportunities during the flights. This system is displayed in figure 1.

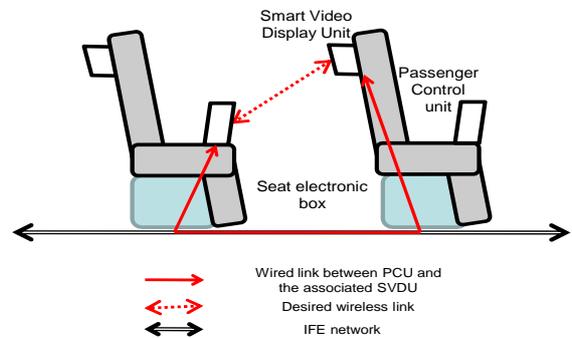

**Figure 1 : Passenger Area Network (PAN)**

This part of the network is the one really close to the passenger. Its purpose is to permit the communication between various elements present in the passenger's seat and to insert these elements in the global IFE system. The system involves the following parts:

- Seat Electronic Box: it is the interface between the IFE network and a small group of passenger's seats. On the passenger side, a Seat Electronic Box can connect 3 seats to the IFE network. It is directly connected to display units.
- Smart Video Display Unit: this unit is the screen placed in the back of the seat in front of the passenger. It allows to interact with the IFE system. This element can be a touch screen one and usually goes with the Passenger Control Unit.

- Passenger Control Units consist of a remote control interacting with the system, and an audio jack connector to plug headphones in order to eventually listen audio contents. Passenger Control Units are required to communicate with Smart Video Display Units, located in the front seat and therefore not connected to the same Seat Electronic Box. Thus, a mean of communication needs to be established between these two boxes.

The main objective is to suppress every specific data wire between these two units (of course power lines are preserved). Consequently, Airlines companies do not need to have rows of seats connected by wires anymore. Since transmission only concerns remote control and audio contents, data rates are rather low (about 3Mbits/s). However this is, for example, still much higher than the data rates allowed in Zigbee protocol [20].

Another application deals with the wireless video streaming, which objective is to suppress each data wire between Seat Electronic Boxes and Smart Video Display Units. It is very similar to the wireless video transmission between a home DVD player and a Full-HD TV. Here, transmission should allow a very high data rate and strong accuracy.

The next two systems, related to aircraft development and maintenance, are located outside the aircraft cabin. As a result, the propagation channel is not indoor anymore and all electronic devices must then be able to withstand to harsh conditions as pressure and temperature.

In the first one, "Structure Health Monitoring" (SHM), sensors distributed inside the structure of an aircraft monitor security relevant parameters, especially in spaces that are difficult to reach for maintenance. A technician can interrogate the sensors either during a maintenance test, in order to check the wearing down of the structure or during a flight in order to warn the pilot in case a problem arises. This results in a medium data rate in the network but a high number of elements connected. In case of upset even leading to data loss, the technician has to reconfigure the system to recover the data.

The most critical application presented here answers to the need of aircraft and satellite manufacturers which desire data describing the plane behavior before it is commercialized. Wireless sensors are spread on the aircraft wings and monitor the pressure around the wing during test flights. This wireless sensor network will enable aerodynamicists to get a large amount of frequently (every few microseconds) updated data points. If one measure gets lost, the whole data set recorded during the flight test becomes unusable. The high data rate transmission therefore has to be extremely accurate. Wireless measurement allows much cheaper and easier deployment than wired sensor measurements.

Figure 2 and table 1 show the data rate versus distance between nodes for the aforementioned applications

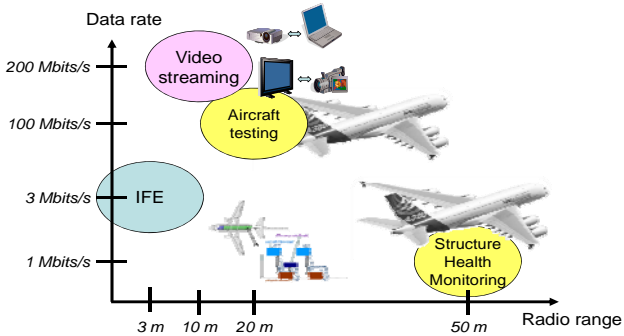

**Figure 2 : Data rate versus radio range for the applications presented**

**Table 1 : specifications for the four applications detailed**

| Appli. | Data rate | Reliability | Range | Nb. of sensor | Channel |
|---|---|---|---|---|---|
| IFE | Medium | Medium | 3-5 m | high | In Cabin |
| Video streaming | Very High | High | 3-10 m | high | In Cabin |
| SHM | Low | Low | 50 m | high | Outside Cabin |
| Aircraft testing | High | High | 10 m | high | Outside Cabin |

### 3. SIMPLE BINARY MODULATION SCHEME

The choice of the modulation scheme has a significant impact on the transceiver design, especially concerning the demodulation circuit. For low power WSN, the transceiver has to remain as simple as possible and therefore, in many applications, only simple binary modulations are considered. These modulation schemes are OOK (On/OFF Keying), FSK (Frequency Shift Keying) or PSK (Phase Shift Keying).

The best tradeoff between bandwidth and energy efficiency is an FSK modulation [1]. However, the drawback is the Bit Error Rate (BER) in FSK, that is higher than the one in Binary – PSK (B-PSK) for a given bit energy per to noise density ($E_b/N_0$).

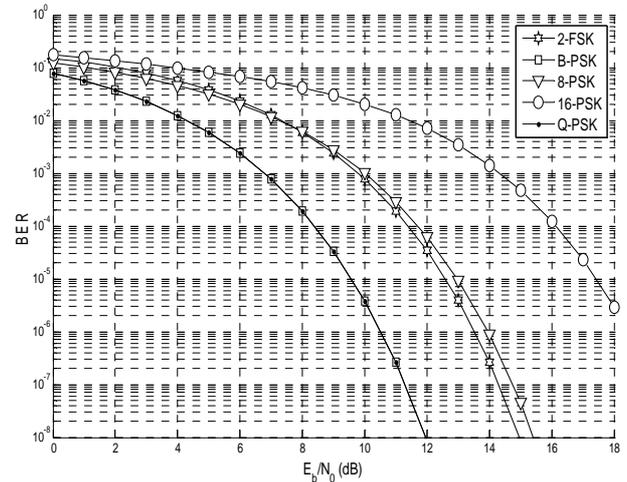

**Figure 3 : Theoretical BER versus Eb/N0 for simple binary modulations in AWGN channel**

Figure 3 shows the required $E_b/N_0$ to achieve a defined BER in a Gaussian channel. B/Q-PSK modulations are obviously more efficient for a given BER, but extremely low available data rates are the main drawback of this modulation scheme. In terms of data rates, B/Q-PSKs are only compatible with such applications as home monitoring and SHM.

As shown in Table 2, the occupied bandwidth for a data rate of 10 MSymbols/s Quaternary-PSK (Q-PSK) link is 13 MHz. In order to achieve very low BER, the $E_b/N_0$, and therefore the radiated power, must be high. In the overcrowded band around 5 GHz, such a WSN will interfere with other systems and consequently not be allowed by regulation's authorities.

In spite of the transceiver's simplicity, this low data rate and high powered modulation cannot be considered as a satisfactory option for WSN radio interface.

**Table 2 : Occupied Bandwidth measured for a raised cosine filtered (roll-off: 0.33) QPSK link**

| Data rate | 500 kS/s | 1 MS/s | 5 MS/s | 10 MS/s |
|---|---|---|---|---|
| Bandwidth at 20dB | 0.66 MHz | 1.3 MHz | 6.5 MHz | 12.9 MHz |

## 4. OFDM

To increase data rate, one can use an architecture simultaneously transmitting multiple signals over a single transmission path. Each signal travels within its own unique frequency range (carrier), which is modulated by the data. Orthogonal frequency-division multiplexing (OFDM) systems break the bandwidth into narrower orthogonal subcarriers and transmit data as parallel streams. Each sub-carrier is modulated with a conventional modulation scheme (such as phase shift keying) at a low symbol rate, maintaining total data rates similar to conventional single-carrier modulation schemes in the same bandwidth.

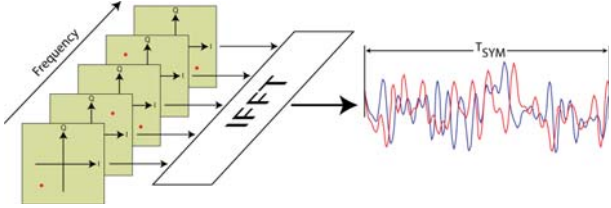

**Figure 4 : IFFT fed with parallels stream allow an Orthogonal Frequency Division Multiplexing**

OFDM uses the Inverse Fast Fourier Transform as a tool to distribute data equally in a certain frequency range (see figure 4). Multiple streams of data feed into an IFFT so that the slower streams are distributed over the bandwidth as individual sub-carriers. A detailed description can be found in [2]. Each input data stream contributes to the time domain waveform, but can be separated by the FFT at the receiver which structure is shown in figure 5.

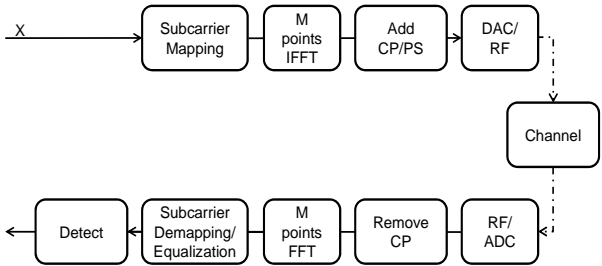

**Figure 5 : OFDM transceiver structure**

Today, this modulation scheme is very popular because it achieves very high data rate and very good spectral efficiency. In a manner similar to IEEE 802.11a/g, an UWB implementation of OFDM was proposed as Multiband OFDM. UWB techniques ensure negligible interferences with existing standards in the 5 GHz band thanks to a -41.3 dBm/MHz limited output power. Multiband OFDM was developed by the IEEE 802.15.3a and standardized by ECMA in [3]. The MB-OFDM physical layer and radio interface is well described in [4] and is summarized in table 3 extracted from [5]. Thanks to its high spectral efficiency and very low output power, MB-OFDM is an interesting option for high Data rate WSN. But, dealing with WSN, long battery lifetime is a key objective, and, as figure 5 shows, the heart of a multiband OFDM transceiver is a 128-point IFFT/FFT which could represent a potential major power drain. Luckily, as FFT is one of the most widespread algorithm in signal processing, several low power FFT/IFFT cores which are suitable for use in WSN are available [6-7].

**Table 3 : Summary of MB-OFDM PHY data rates**

| Data Rate (Mbps) | Modulation Type | Coding Rate (R) | Frequency Repetition Factor | Time Repetition Factor |
|---|---|---|---|---|
| 53.3 | QPSK | 1/3 | 2 | 2 |
| 80 | QPSK | 1/2 | 2 | 2 |
| 106.6 | QPSK | 1/3 | 1 | 2 |
| 160 | QPSK | 1/2 | 1 | 2 |
| 200 | QPSK | 5/8 | 1 | 2 |
| 320 | DCM | 1/2 | 1 | 1 |
| 400 | DCM | 5/8 | 1 | 1 |
| 480 | DCM | 3/4 | 1 | 1 |

Since no other element seems to be a big power consumer, we can draw a state of the art of an OFDM transmitter. Some of the most recent data are summarized in table 4. Three different kinds of developments can be found in literature:

- Receiver front end for, which includes LNA, Mixer, filter, VGA and Local Oscillator (QVCO, frequency divider and synthesizer).
- Complete transceiver front end, which includes a receiver front end, a transmitter front end and a power amplifier.
- Full transceiver development, which includes the front end, FFT/IFFT, DAC/ADC, modulator and coding.

Leaving [15] aside, because of missing data, the best development is in CMOS 65 nm and consumes around 200 mW, which is low compared to the 1W with 802.11b types transceivers needs. But even when all elements are optimized, the consumption level is too high to enable the use of a power limited device. Figure 6 shows that the power amplifier is the most consuming block independently of the optimization.

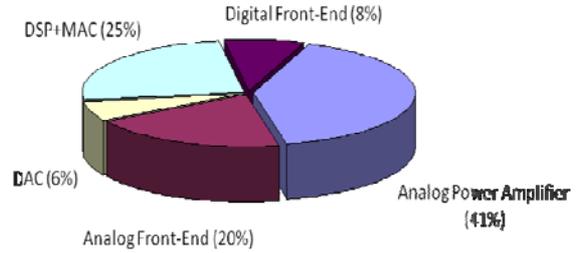

**Figure 6 : OFDM transmitter power consumption budget**

This is due to the high Peak Average Power Ratio (PAPR), one of the major drawbacks of ODFM for low power application. Indeed, OFDM symbols are a combination of a great number of subcarriers. Consequently, subcarrier voltages can be added in-phase at some point within the symbol, resulting in very high instantaneous power as detailed in figure 7. If an OFDM signal has 128 carriers, like MB-OFDM, each with maximum amplitude of 1mV, we could get a 128mV maximum amplitude in the signal. Considering the 128 carriers each with a normalized power of 1mW, when all 128 carriers combine at their maximum point, the power can be as large as 21 dBm. Such a peak is much higher than the average power.

As a result, OFDM signals require highly linear power amplifiers to avoid excessive inter modulation distortion. To achieve this linearity, the amplifiers have to be designed with a large back off from their peak power. This results in a very poor power efficiency, which is not compatible with the power consumption reduction and can turn out to be a real problem for OFDM WSN implementation.

**Table 4 : State of the art for MB-OFDM transceivers**

| Ref. | [8] | [9] | [10] | [11] | [12] | [13] | [14] | [15] |
|---|---|---|---|---|---|---|---|---|
| Year | 2005 | 2005 | 2006 | 2006 | 2006 | 2006 | 2007 | 2005 |
| standard | ECMA MB-OFDM 3.1–10 GHz | ECMA MB-OFDM 3.1–8.2 GHz | ECMA MB-OFDM 3.1–8 GHz | ECMA MB-OFDM 3.1–8 GHz | ECMA MB-OFDM 3.1–4.8 GHz | ECMA MB-OFDM 3.1–9.5 GHz | ECMA MB-OFDM 3.1–4.8 GHz 6 – 8.1 GHz | ECMA MB-OFDM 3.1–4.8 GHz |
| Type | Front END | Front END | Front END | Front END | FULL Transceiver | Transceiver Front END | Transceiver Front END | Transceiver Front END |
| Voltage | 2.7 V | 2.7 V | 1.5 V | 2.3 V | 3.3 V/1.5 V | 1.1 V | 1.2 V | 1.5 V |
| Rx current | 47 mA 127 mW | 22 mA 59 mW | 22.5 mA 34 mW | 19.5 mA 45 mW | 100 mA | 161 mA 177 mW | 43 mA 52 mW | 40 mA 60 mW |
| Tx current | | | | | 70 mA | 76.4 mA 84 mW | 38 mA 46 mW | |
| Synthesizer | 27 mA 73 mW | 66 mA 179 mW | 59 mA 89 mW | N/A | N/A | 43 mA 47 mW | LO : 52 mA Without PLL 62 mW | LO : 30 mA 45 mW with PLL |
| Total power RX/TX | 200 mW | 238 mW | 123 mW | N/A | N/A | 224 mW / 131 mW | 114 mW / 108 mW | 105 mW |
| Process | QUBiC4G SiGe | 0.18 um SiGe BICMOS | 0.18 um CMOS | 0.18 um CMOS | 0.13 um CMOS | 90 nm CMOS | 65 nm CMOS | 65 nm CMOS |

Different PAPR reducing methods were developed. The most important are:
- *Signal clipping* at a desired power level. Matlab simulation could be done with a 16 QAM modulation, 64 points for IFFT and a clipping such as the squared signal divided by the variance stay beyond 10. A 56 dB PAPR before clipping and a 14 dB PAPR after the clipping is measured. The reduction is important but it introduces other distortions and Inter Carrier Interferences.
- *Selective mapping*. The signal is multiplied by a set of codes; an IFFT is done on each and the one with the least PAPR is picked. Codes are similar to CDMA. But these techniques impose parallel operations and the selection of the best one. In power constraint application, this method exhibits a too large complexity.
- *Partial FFT*. The signal is divided into M equal length disjoint signal sub-blocks. Then, each block is fed to an Inverse Fast Fourier Transform (IFFT) of length N and multiplied by a pure rotation weighting factors, and then they are combined (added) to constitute the transmitted signal [16].

These improvement techniques are well suited for those applications without strong power consumption constraints. They can be applied in such applications as IFE, but not for structure health monitoring or for collecting data describing aircrafts' behavior. The use of MB-OFDM techniques for WSN is very application dependant and requires other ultra low power WSN radio interface techniques.

## 5. SC-FDMA

SC-FDMA ([17]) is an option to reduce OFDM systems' PAPR and to increase self powered systems' lifetime. This modulation scheme is not just an improvement of OFDM but a new modulation scheme. SC-FDMA which utilizes single carrier modulation and frequency domain equalization has similar performance and basically the same overall complexity as OFDM systems. But, one significant advantage over OFDM is that the SC-FDMA signal has lower PAPR because of its inherent single carrier structure.

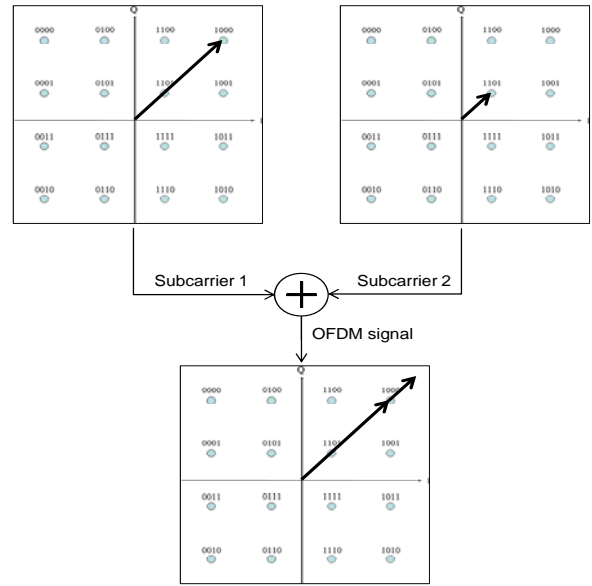

**Figure 7 : The power of the OFDM signal could be the sum of the maximum power of each subcarrier.**

It is currently adopted as the uplink modulation and multiple access scheme in 3GPP Long Term Evolution (LTE), or Evolved UTRA [18], here lower PAPR greatly affects the handheld terminal in terms of transmit power efficiency and manufacturing costs. This property can turn out to be very interesting in the development of ultra low power wireless sensor network application. SC-FDMA is an extension of Single Carrier modulation with Frequency Domain Equalization techniques (SC-FDE). SC-FDE is the frequency domain equivalent of what is done in a conventional linear time domain equalizer. For broadband channels, conventional time domain equalizers are impractical because of the complexity. Equalizers have to use a great number of coefficients and have to perform very high speed operations. OFDM is very efficient because the equalization is done in the frequency domain by a simple product of the signal with the inverse channel coefficient. Figure 8 shows block diagrams of SC-FDE, SC-FDMA and OFDM systems.

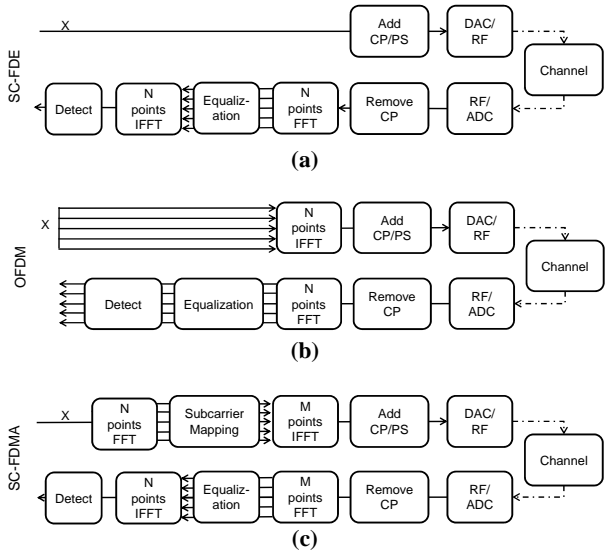

**Figure 8 : Transmitter structures of SC/FDE (a), OFDM (b) and SC-FDMA (c)**

Comparing SC-FDE and OFDM architectures, it is obvious that they feature some similarities. Both use the same component blocks. This similarity leads to the fact that the two systems feature the same spectral efficiency. The only difference is the location of the IFFT block. But SC-FDE architecture, besides the advantages of having very low fluctuations as shown in figure 9, features a lower sensitivity to carrier frequency offset and lower complexity at the transmitter level.

SC-FDMA architecture is an extension of SC-FDE to accommodate multi user access. As in OFDM, the transmitter in a SC-FDMA (or SC-FDE) system use different orthogonal subcarriers to transmit information symbol. But the only difference is that they are transmitting sequentially, rather than in parallel. All subcarrier cannot be added in phase, so this arrangement reduces considerably the envelope fluctuation.

First, the transmitter of an SC-FDMA system groups the modulation symbols into blocks each containing $N$ symbols. Next it performs an $N$-point FFT to produce a frequency domain representation of the input symbols. It then maps each of the $N$-FFT outputs to one of the $M$ ($> N$) orthogonal subcarriers that can be transmitted. If $N = M/Q$ and all terminals transmit $N$ symbols per block, the system can handle $Q$ simultaneous transmissions without co-channel interference. $Q$ is the bandwidth expansion factor of the symbol sequence. As in OFDM, a $M$-point IFFT transforms the subcarrier amplitudes to a complex time domain signal.

The receiver transforms the received signal into the frequency domain via a FFT, de-maps the subcarriers, and then performs frequency domain equalization. Most of the well-known time domain equalization techniques, such as minimum mean square error (MMSE) equalization can be applied to the frequency domain equalization.

In SC-FDMA systems, there are two methods to choose the subcarriers for transmission:
- *Localized subcarrier mapping* (LFDMA). Each sensor uses a set of adjacent subcarriers to transmit its symbols.
- *Distributed subcarrier mapping* (DFDMA). The subcarriers used by a sensor are spread over the entire signal band. One realization of DFDMA is Interleaved FDMA (IFDMA) where occupied subcarriers are equidistant to each other.

Simulation results presented in [19] lead to the conclusion that IFDMA is more desirable than LFDMA in terms of PAPR and power efficiency. PAPR performance is superior by 4 to 7 dB to that of LFDMA. When we consider a pulse shaping to control adjacent channel interference, we find a narrower difference between LFDMA and IFDMA in terms of PAPR performance.

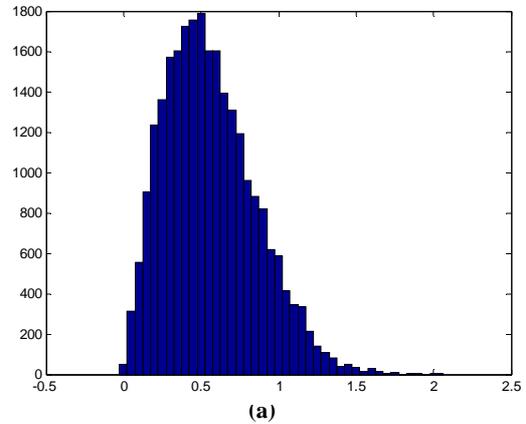

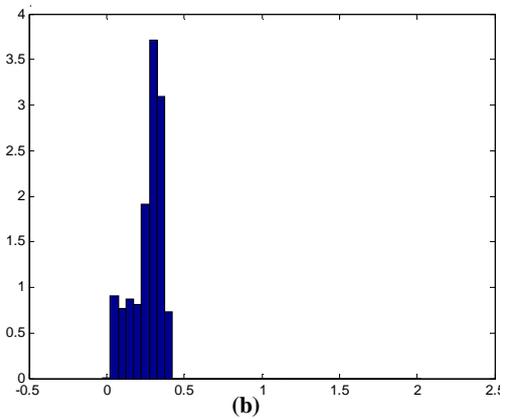

**Figure 9 : (a) Histograms of an OFDM signal at the entry of the amplifier.**
**(b) Histogram of a SC-FDE signal at the entry of the amplifier.**

So, in an ultra low power application, single carrier modulation offers a good alternative to OFDM. By using SC-FDE or its application for multiuser SC-FDMA, we are able to develop low power, high data rate and high performance wireless sensor networks. The scheduling should use IFDMA in order to have better PAPR.

## 6. DESIGN METHODOLOGY

Single carrier modulation offers the best option to develop low power and high data rate transceivers. However, no realization with real measured power consumption of such a transceiver can be found in literature. In order to realize such a modulator and to compare it with OFDM realizations, we chose an appropriate design methodology.

The conception flow is presented in figure 10. The first step is a Matlab/Simulink model of a full transceiver for either OFDM or SC-FDMA modulation scheme. When the model of the simple transceiver is validated, different tests can be performed, for example, on synchronization techniques or frequency allocations and the best solutions can be selected easily. Figure 11 shows a simplified Simulink model for an OFDM transceiver. This model allows the realization of a compact OFDM transceiver which can be used to evaluate the usefulness of redundant elements like, for example, cyclic prefix. With model based design, we are able to build the simplest transceiver model for a specific need. This optimal model will be implemented with Simplify DSP IP on a Simulink model in order to generate VHDL code. Thus, we have a FPGA development of an OFDM and a SC-FDMA transceiver which are tested to check the spectral efficiency, the data rate and the overall performance. We can have a link between model and hardware realization. The gap and the performance of the Simplify DSP tools between our realization and the state of the art can be evaluated. At each step, we are able to come back to the system model and check the accuracy of it and to test new approaches to solve problems that appear in hardware realization. The design flow allows the quick realization of both transceivers.

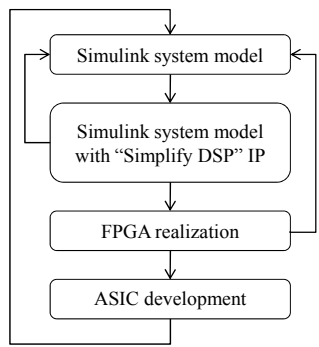

**Figure 10 : Design methodology**

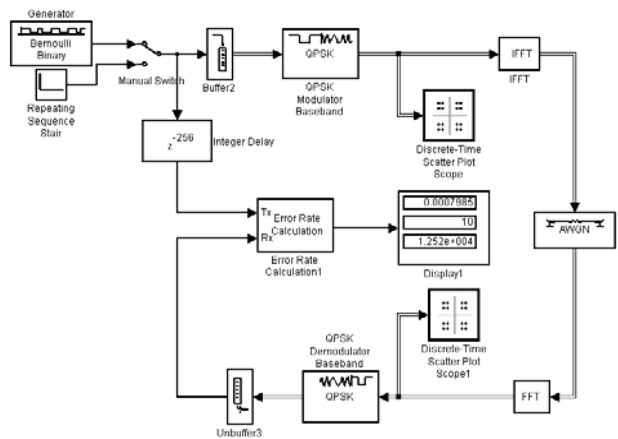

**Figure 11 : Simple Simulink model for OFDM transceiver**

### 7. CONCLUSION

In this paper, we have demonstrated that the biggest problem caused by OFDM modulation for low power applications is its high PAPR. As Single Carrier approaches have the same complexity and performance, they are well suited alternative in the realization of high data rate Wireless Sensor Networks
We have proposed an original design methodology where, we can implement a SC-FDMA transceiver for WSNs use on FPGAs.